\documentstyle[aps,prb,multicol,epsf,picinpar]{revtex} 

\def\ep {\epsilon}

\def\e2 {\epsilon-\epsilon_k}
\def\be {\begin{equation}}
\def\ee {\end{equation}}
\def\bea {\begin{eqnarray}}
\def\eea {\end{eqnarray}}
\def\om {\omega}
\def\rs {\tau_{sp}^{-1}}

\begin{document}
\draft           
\draft
\title{Dephasing from interactions and spin disorder }

\author{ George Kastrinakis}

\address{
Institute of Electronic Structure and Laser (IESL), Foundation for
Research and Technology - Hellas (FORTH), 
P.O. Box 1527, Iraklio, Crete 71110, Greece$^*$}

\date{October 10, 2002}

\maketitle

\begin{abstract} 

We calculate the dephasing rate of the electrons in the presence 
of interactions and elastic spin disorder scattering. 
In the frame of a self-consistent diagrammatic treatment,
we obtain saturation of the dephasing rate in the limit of zero temperature
for spin-orbit disorder in 2 dimensions. 
This result is in agreement with relevant experiments.
\end{abstract}

\vspace{.3cm}

The dephasing rate $\tau_\phi^{-1}$ provides a measure of the loss of 
coherence of
the carriers. The saturation of the dephasing rate at low temperature seen
in {\em some} samples 
\cite{mo1,mo2,nat,mar,piv,gou,lin,lin2,pie} has attracted a vigorous interest.

In the presence of spin-less disorder, the Cooperon 
(particle-particle diffusion
correlator) is given by
\be
C_0(q,\om) = \frac{1}{2\pi N_F \tau^2}
\frac{1}{Dq^2 - i \om + \tau_{\phi}^{-1}} \;\;.
\ee
$D$ is the diffusion coefficient,
$N_F$ is the density of states at the Fermi level and $\tau^{-1}$ 
the total impurity scattering rate. We will work in the diffusive regime
$\epsilon_F \tau >1$ ($\hbar=1$), $\epsilon_F$ being the Fermi energy.

Previous studies \cite{aak,fa,aag,aav,vk1,vk2,gz1,gzs,gz2,ifs,zvr} 
have focused on the calculation
of $\tau_{\phi}^{-1}$ in the absence of spin-scattering disorder.
Here we calculate $\tau_{\phi}^{-1}$ 
in the presence of spin-scattering disorder, whereupon the Cooperon 
becomes spin-dependent. The relevant terms $C_i$ are shown in fig. 1.
We start by giving the explicit form of these $C_{1,2}^o$ {\em without} 
a dephasing rate. 
For the case $\tau_{so}^{-1}>0, \tau_S^{-1}=0$ - with $\tau_{so}^{-1}$ the
spin-orbit impurity scattering rate and $\tau_S^{-1}$ the magnetic impurity
scattering rate - they are given by \cite{berg}
\be
C_{1,2}^o = b_{1,2} \Big[ \frac{1}{Dq^2 -i \om}
-\frac{1}{Dq^2 -i \om + 3/(4 \tau_{so})} \Big] \;\;, 
\ee
with 
$b_1=(3\tau_{so}/\tau -2)/(2u), b_2=-1/u, u=4\pi N_F \tau^2$.
We should emphasize that the impurity scattering considered is elastic.

To calculate the dephasing rate, 
we solve the appropriate Bethe-Salpeter equations for {\em all} three
Cooperons $C_i(q,\om)$, $i=0,1,2$.  These are shown schematically
in fig. 2:
\bea
C_0 = C_0^o + C_0^o W C_0 \;\;,  \label{exsc1}\\
C_1 = C_1^o + C_1^o [F C_1+ R C_2] + C_2^o [R C_1 + F C_2] \;\;, \\
C_2 = C_2^o + C_2^o [F C_1+ R C_2] + C_1^o [R C_1 + F C_2] \;\;. 
\label{exsc2}
\eea
In fig. 3 we show explicitly the components of the 
self-energy terms $\Sigma=W, F, R$.
These quantities are given by
\bea
W =  \Sigma_0 + d_0 \Sigma_1 \;\;, F= \Sigma_1 + d_1 \Sigma_0 \;\;,
R = d_2 \Sigma_2 \;\;.
\eea
Here, 
$d_0=d_1= \{1/(\ep_F\tau) + 4\pi\}/(2 \pi^2 \ep_F \tau_{so})$ and 
$d_2=-2 \{1/(\ep_F\tau_{so}) + 2 \pi\tau/\tau_{so}\}/\pi$.
The terms containing the factors $d_i$ - with a spin impurity line either
looping
around the Cooperon or crossing it - provide the coupling between
the spin-independent and the spin-dependent Cooperons, and they turn
out to be {\em crucial} for the saturation of the dephasing rate obtained
below.
In the spirit of ref. \cite{fa} we obtain
\be
\Sigma_i \equiv \Sigma_i(q=0,\omega=0)\simeq \\
- \sum_q \int_{-\infty}^{\infty} dx
\frac{ C_i(q,x+i 0) \; \text {Im} V(q,x+i 0)}{\sinh(x/T)} \;\;,
\ee
where
\be
V(q,\om)=\frac{v_q}{1+v_q \Pi(q,\om)} \;\;, \;\; \\
\Pi(q,\om) = \frac{N_F}{2} \Big[ \frac{Dq^2}{Dq^2 - i\om} +
\frac{Dq^2 + \rs}{Dq^2 - i\om + \rs} \Big] \;\;. \label{pot}
\ee
Here $v_q=2\pi e^2/q$ is the bare Coulomb interaction 
and $\rs = (4/3)(\tau_S^{-1} + \tau_{so}^{-1})$ the total spin scattering rate.
We make the approximation \cite{fa} 
$\int_{-\infty}^{\infty} dx F(x)/\sinh(x/T)
\simeq T\int_{-T}^T dx F(x)/x$. 
The bosonic modes with energy greater than $T$ manifestly do {\em not}
contribute to the self-energy and the dephasing process.

Since the coefficients $d_i\ll 1$ for $\ep_F \tau_{so} \gg 1$,
we can decouple a
$2\times 2$ system of equations involving only $C_{1,2}$, to facilitate the
solution of eqs. (\ref{exsc1}-\ref{exsc2}).
Subsequently, $C_{1,2}$ are given by
\be
C_{1,2}(q,\om) = S_{1,2} \Big[ \frac{1}{Dq^2 -i \om + r_-} - 
\frac{1}{Dq^2 -i \om + r_+} \Big] \;\;.
\ee
Here 
$r_+=(b_1-b_2)(\Sigma_2-\Sigma_1), r_-=-(b_1+b_2)(\Sigma_2+\Sigma_1)$,
$S_i=a_{o}\{b_i (Dq^2 -i \om) + M_i\}/X,
X=2 (b_1 \Sigma_2 + b_2 \Sigma_1)$ and 
$a_o=8 N_F \tau^4/ D $.

The self-energies are given by 
\be
\Sigma_i = \frac{T a_o}{X} \Big\{(M_i-b_i r_-) 
\ln{\Big(\frac{z_o-r_-}{T-r_-}\Big)}
- (M_i-b_i r_+) \ln{\Big(\frac{z_o-r_+}{T-r_+}\Big)} \Big\} \;\;, 
\label{exsi}
\ee
with $i=1,2$, $z_o = \tau^{-1}$ 
and $M_1=(b_2^2-b_1^2) \Sigma_1, M_2=-(b_2^2-b_1^2) \Sigma_2$.

Taking 
\be
b_1=-b_2 \;\; \;(\tau_{so}=\frac{4}{3} \tau) \;\;,
\ee
we obtain a simplified version of these equations:
\be
\Sigma_1 = T a_o b_1 
\ln{\Big(\frac{z_o +4 b_1 \Sigma_1}{T +4 b_1\Sigma_1}\Big)}
\;\;,
\ee
and $\Sigma_2=-\Sigma_1$.
In the limit $T\rightarrow 0$ we obtain the solution
\be
\Sigma_1 = - z_o/(4 b_1) + O(e^{-1/T}) \;\;\;.
\ee 
This finite solution for $\Sigma_{1,2}$ for $T\rightarrow 0$ is 
a generic fact for any finite value of $\tau_{so}^{-1}$. \cite{com1}

In 1-D we obtain similarly
\be
\Sigma_i = \frac{T a_{1D}}{X} \Big\{
\frac{M_i-b_i r_-}{\sqrt{r_-}}\Big[ f(z_o,r_-) - f(T,r_-) \Big] \\
-\frac{M_i-b_i r_+}{\sqrt{r_+}}\Big[ f(z_o,r_+) - f(T,r_+) \Big] 
\Big\}\;\;,
\ee
with $f(x,y)=\ln\{(\sqrt{x}-\sqrt{y})/(\sqrt{x}+\sqrt{y})\}$ and
$a_{1D}=2\pi \sqrt{D} a_o$.
In 3-D we have
\bea
\Sigma_i = \frac{T a_{3D}}{X} \Big\{ b_i X \sqrt{z_o} +
(M_i-b_i r_-)\sqrt{r_-}\Big[ f(z_o,r_-) - f(T,r_-) \Big]
-(M_i-b_i r_+)\sqrt{r_+}\Big[ f(z_o,r_+) - f(T,r_+) \Big]
\Big\}\;\;, \nonumber
\eea
with $a_{3D} = 2 a_o/\pi$.
Both the 1-D and 3-D cases differ from 2-D in that no {\em finite}
solution exists for $\Sigma_{1,2}$ in the limit $T\rightarrow 0$.
This is due to the presence of the factors 
$\sqrt{r_\pm} \propto \sqrt{\Sigma_i}$
in the right-hand side of the respective eqs. for $\Sigma_i$.

It turns out that 
\be
\tau_{\phi}^{-1} \simeq  - b_0 d_0 \Sigma_1 
= \frac{2 [4\pi + 1/(\ep_F \tau)]}{\pi^2 \tau (\ep_F \tau_{so}) 
(3 \tau_{so}/\tau -2)} \;\; \;\;, \label{ext}
\ee
with $b_0=1/(2 \pi N_F \tau^2)$.
Hence $\tau_{\phi}^{-1}$ {\em saturates} in the limit $T\rightarrow 0$
in 2-D. This is to be contrasted with the absence of spin disorder,
where it has been shown that $\tau_{\phi}^{-1} \propto T \ln{T}
\rightarrow 0$ \cite{aak,fa,aag,aav,vk1,vk2}, 
and this fact yields eq. (\ref{ext}).
Moreover, we would like to emphasize that other processes in $\Sigma$, 
which are first order in the interaction $V(q,\omega)$, e.g. with $V$ 
crossing diagonally the Cooperon, do not modify qualitatively the result
in eq. (\ref{ext}).

Now, the total correction to the conductivity can be written as
\be
\delta \sigma_{tot} = \delta \sigma_o + \delta \sigma_{sp} \;\;,
\ee
where the first term is the spin independent one  - involving $C_0$ -
and the second term the spin dependent one - involving $C_{1,2}$. 
The latter is expected to saturate always in the low $T$ limit.
The form usually fit to experiments \cite{mo1,nat} for the former is
\be
\delta \sigma_o = \sigma_0 \Big[ \ln(y) - \Psi(y + \frac{1}{2})\Big] \;\;,
\ee
where $\Psi$ is the digamma function, $y=1/4 e H D \tau_\phi$, $H$
the magnetic field and $\sigma_0 = e^2/(2 \pi^2)$.
Obviously, this expression saturates for a saturating $\tau_{\phi}^{-1}$,
and vice-versa for the experimental determination of $\tau_{\phi}^{-1}$.

A number of experiments 
show saturation of $\tau_\phi^{-1}$ in the zero temperature limit.
The samples in which saturation is observed are made of
elements with a high atomic number, which induces a substantial 
spin-orbit scattering. Moreover, several of the samples, in which the 
dephasing saturation is observed, are truly 2-dimensional, such as the wires
in refs. \cite{mo1,mo2,nat}, the quantum dots in refs. \cite{mar,piv} etc.
We believe that the observed saturation
can be understood in the frame of our results above.

\vspace{.3cm}
I have enjoyed useful discussions/correspondence with Peter Kopietz and 
Pritiraj Mohanty.

\vspace{.3cm}
$^*$ e-mail : kast@iesl.forth.gr

\vspace{4cm}

\begin{figure}
\begin{center}
\epsfxsize8cm
\epsfbox{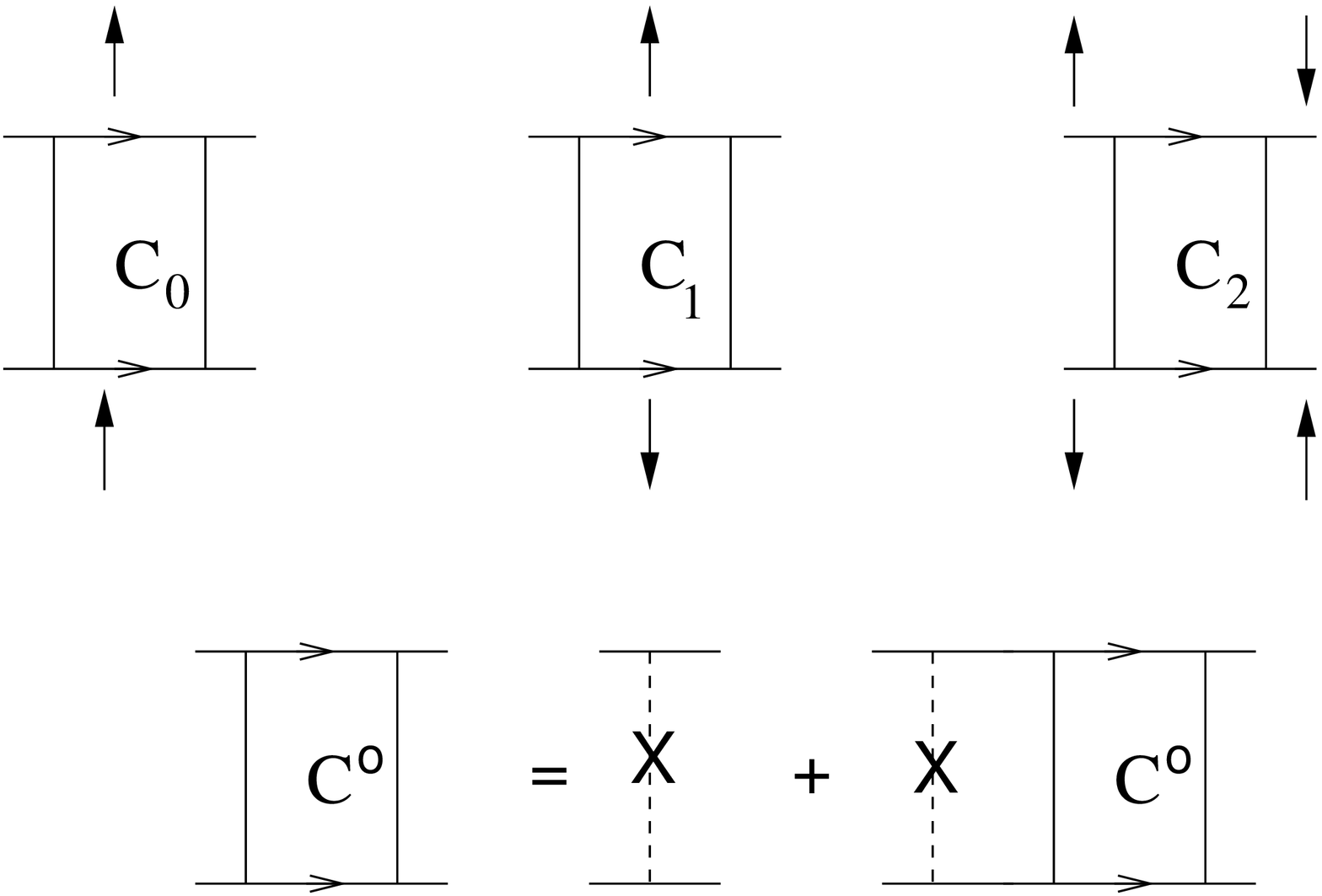}
\vspace{.1cm}
\centerline{Fig. 1}
\end{center}
The 3 Cooperons $C_0,C_1,C_2$. Note the spin indices. The Cooperons $C_i^0$
do not contain a dephasing rate. The dashed line with the cross stands
for impurity (disorder) scattering.
\end{figure}

\begin{figure}
\begin{center}
\epsfxsize10cm
\epsfbox{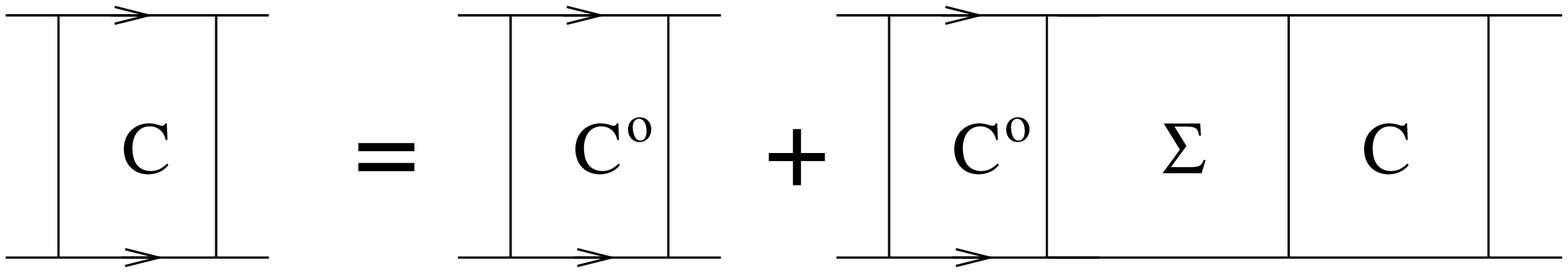}
\vspace{.1cm}
\centerline{Fig. 2}
\end{center}
Schematic form of the equations (\ref{exsc1}-\ref{exsc2}) involving the
Cooperons and the self-energies $\Sigma$.
\end{figure}

\begin{figure}
\begin{center}
\epsfxsize6cm
\epsfbox{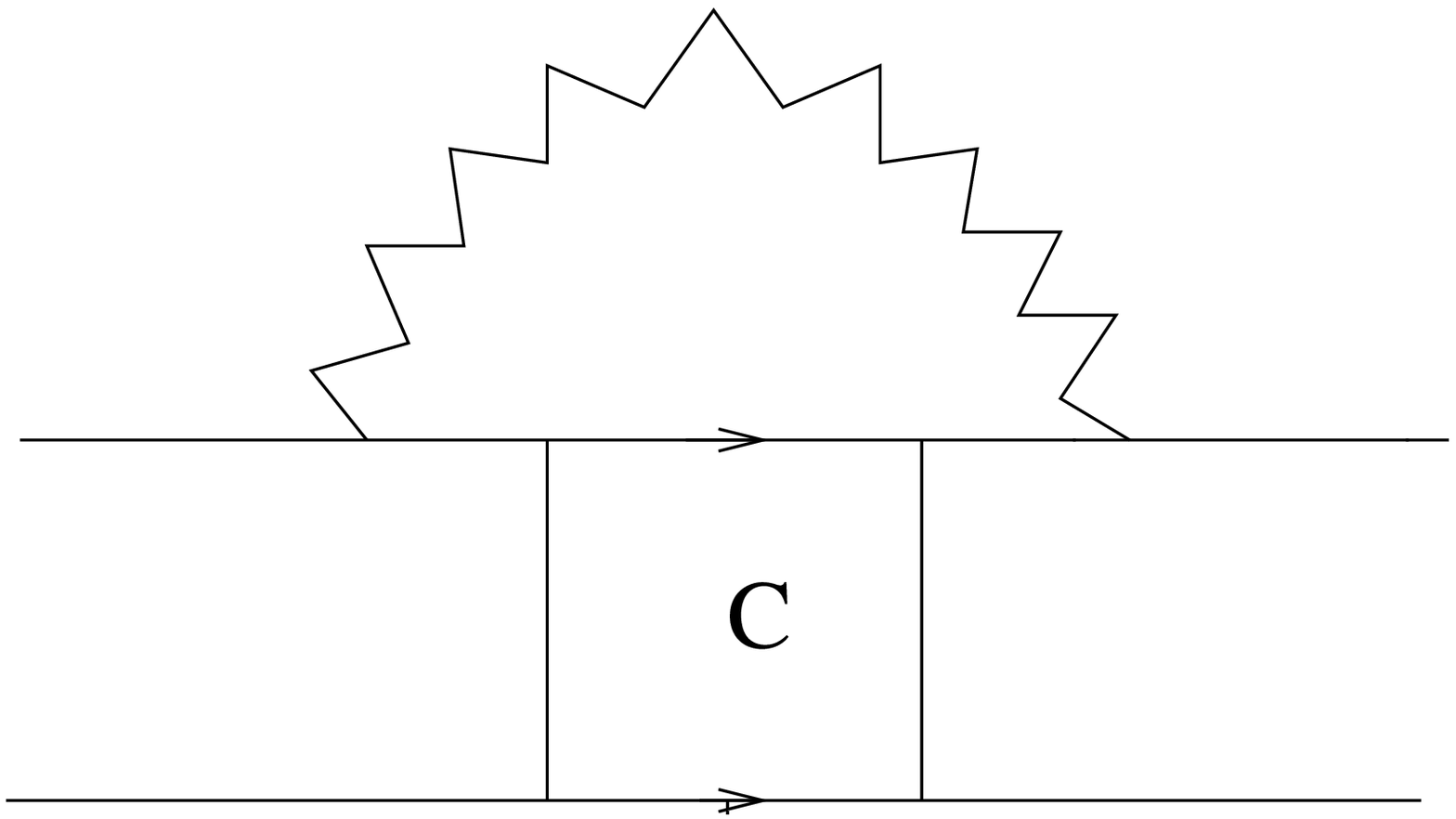}
\epsfxsize6cm
\epsfbox{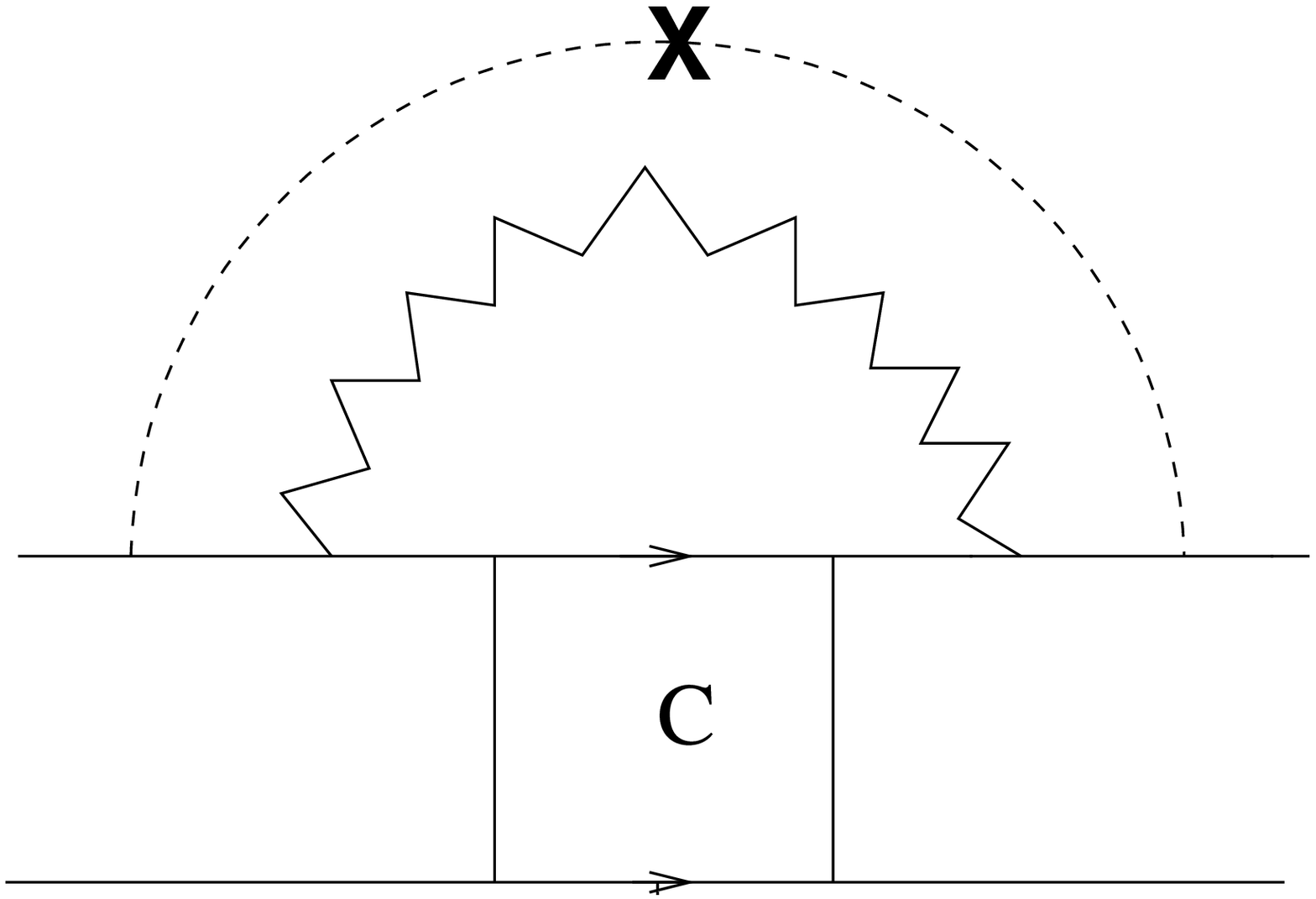}
\vspace{.3cm}
\epsfxsize6cm
\epsfbox{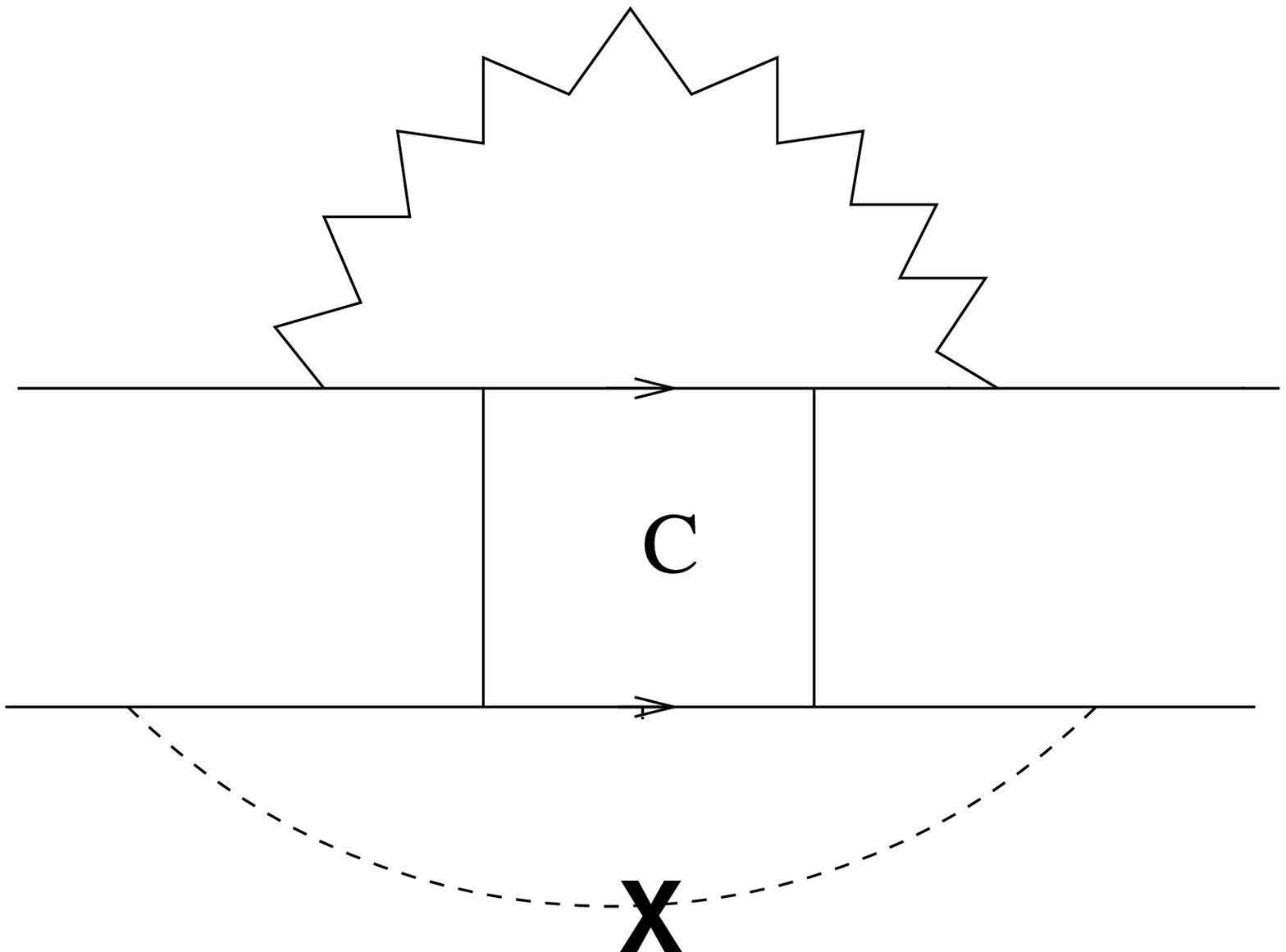}
\vspace{.3cm}
\epsfxsize6cm
\epsfbox{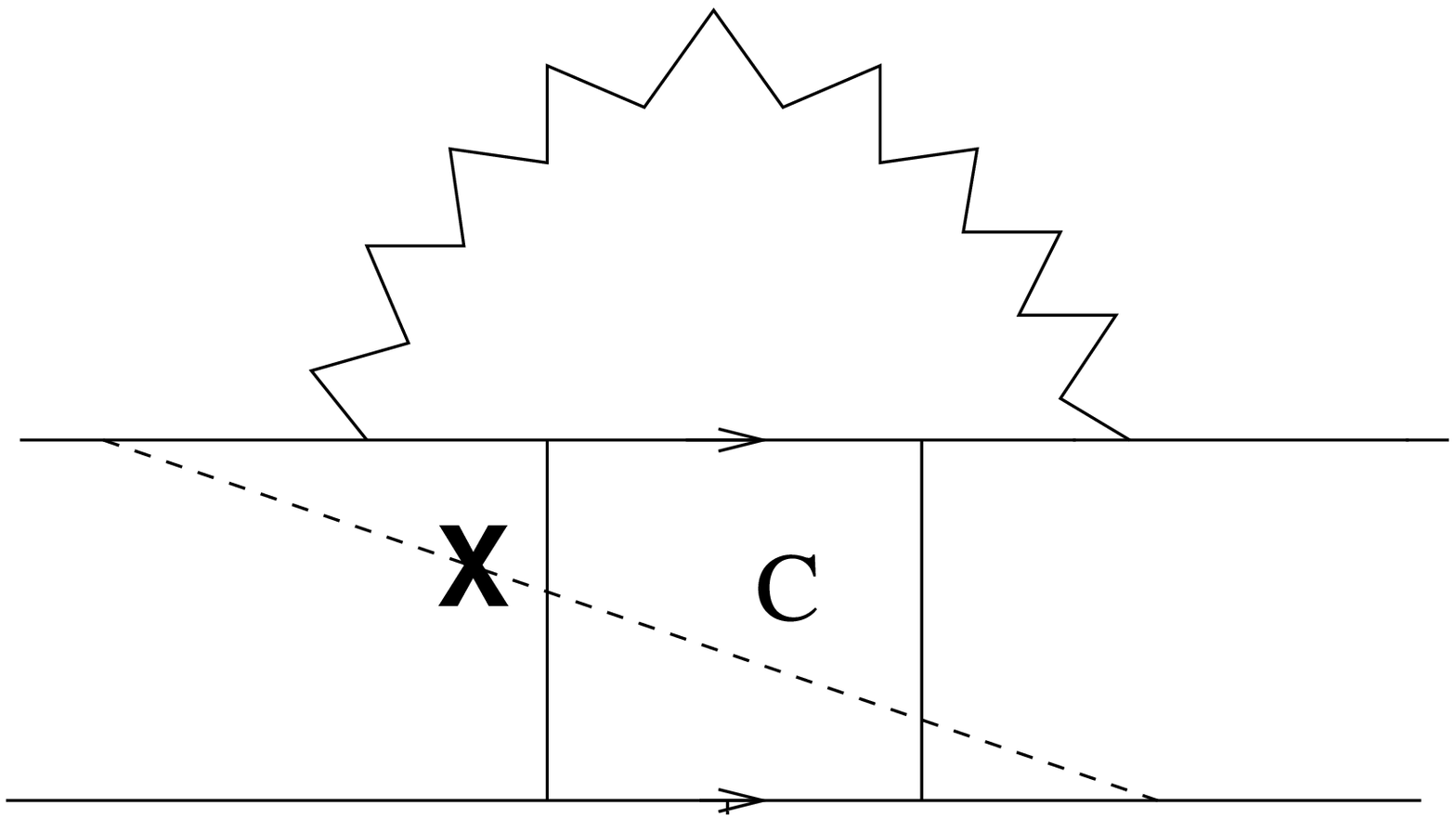}
\centerline{Fig. 3}
\end{center}
The various components of the self-energy. The wiggly line represents
the screened Coulomb interaction of eq. (\ref{pot}). The diagrams with
the extra spin-disorder impurity line are crucial for the saturation
of $\tau_{\phi}^{-1}(T\rightarrow 0)$. We consider {\em all} possible
variations of the diagrams shown here.
\end{figure}


\begin{references}
\bibitem{com1}
Note that $\Sigma_1 \propto T\ln{(z_o/T)} \rightarrow 0$ is
also a solution, yielding however $C_{1,2}(q,\omega)\rightarrow 0$
identically.
\end{references}
\end{document}